\documentstyle[12pt,amssymb,epsfig]{article}
\begin{document}

\title{Lyapunov exponent in quantum mechanics. A phase-space approach.}
\author{{\it V. I. Man'ko}{\thanks{%
on leave from the P. N. Lebedev Physical Institute, Moscow, Russia}\quad and 
}{\it R.Vilela Mendes}\thanks{%
corresponding author: vilela@alf1.cii.fc.ul.pt} \\
{\small Grupo de F\'{i}sica--Matem\'{a}tica,}\\
{\small \ \ Complexo Interdisciplinar, Universidade de Lisboa}\\
{\small \ \ Av. Prof. Gama Pinto, 2, 1699 Lisboa Codex, Portugal}}
\date{}
\maketitle

\begin{abstract}
Using the symplectic tomography map, both for the probability distributions
in classical phase space and for the Wigner functions of its quantum
counterpart, we discuss a notion of Lyapunov exponent for quantum dynamics.
Because the marginal distributions, obtained by the tomography map, are
always well defined probabilities, the correspondence between classical and
quantum notions is very clear. Then we also obtain the corresponding
expressions in Hilbert space.

Some examples are worked out. Classical and quantum exponents are seen to
coincide for local and non-local time-dependent quadratic potentials. For
non-quadratic potentials classical and quantum exponents are different and
some insight is obtained on the taming effect of quantum mechanics on
classical chaos. A detailed analysis is made for the standard map.

Providing an unambiguous extension of the notion of Lyapunov exponent to
quantum mechnics, the method that is developed is also computationally
efficient in obtaining analytical results for the Lyapunov exponent, both
classical and quantum.
\end{abstract}

\section{Introduction}

Classical chaotic motion is characterized by the existence of positive
Lyapunov exponents or positive Kolmogoroff-Sinai entropy. Because the
definition of these quantities is based on the properties of classical
trajectories in phase space, it is not obvious what the corresponding
quantities in quantum mechanics should be. This situation led to the
proposal of many different quantities to characterize the quantum behavior
of classically chaotic systems (\cite{Connes} - \cite{Winter}). To be sure
that one constructs quantum mechanical functionals, with exactly the same
physical meaning as the classical quantities which characterize classical
chaos, the natural suggestion would be to use also a phase-space formulation
for quantum mechanics, rather than the usual Hilbert space formulation. The
difficulty here lies in the fact that quantum phase-space is a
non-commutative manifold with the usual pointwise product of functions being
replaced by the $*$-product\cite{Lichne1}. One possible solution is to use
the tools of non-commutative geometry for this formulation. Another
approach, however, is to look for (commutative) quantities which have the
same formal structure both in classical and in quantum mechanics. The Wigner
function\cite{Wigner32}, which some authors have attempted to use for this
purpose, is not the appropriate choice because, unlike the classical
probability distributions, it is not positive definite. In fact, the Wigner
function is only correctly interpreted in a non-commutative geometry setting
or, alternatively, as an operator symbol which by the Weyl map corresponds
to a well defined operator in Hilbert space\cite{Lichne1}.

There is however a set of phase-space quantities that have the same
mathematical nature in both classical and quantum mechanics. This is the set
of marginal distributions of the symplectic tomography formulation (\cite
{Mancini95} - \cite{elaf98}), which are in both cases well-defined
probability distributions. In Section 2, we review the symplectic tomography
formulation of classical and quantum mechanics. In both cases the dynamics
is defined by a set of marginal probability distributions. The difference
between classical and quantum mechanics comes only on the modification of
the equations of motion, which in this formulation is the analog of the
Moyal deformation\cite{Moyal49} \cite{Lichne1} of the phase-space algebra.

Once the appropriate phase-space quantities are identified and classical
Lyapunov exponents (and local entropies) are formulated in terms of
probability distributions, the transition to quantum mechanics is rather
straightforward.. This is discussed in Section 3. Having obtained the
Lyapunov exponent for quantum mechanics from the marginal probability
distributions, one is then also able to obtain the corresponding Hilbert
space expression.

Some examples are then computed, involving kicked systems on the line, on
the 2-torus and on the circle. Classical and quantum exponents are seen to
coincide for time-dependent local and nonlocal quadratic potentials. For
non-quadratic potentials classical and quantum exponents are different. A
characterization is obtained for the origin of the taming effect of quantum
mechanics on classical chaos in the standard map. This is a step towards a
rigorous characterization of this effect because it refers, as in the
classical case, to the behavior of the Lyapunov exponents and not to
indirect quantities like the energy growth.

\section{Symplectic tomography of classical and quantum states}

\subsection{Classical mechanics}

States in classical statistical mechanics are described by a function $\rho
\left( q,\,p\right) $, the probability distribution function in $2n$%
-dimensional phase space $\left( q\in R^{n},p\in R^{n}\right) $, with
properties 
\[
\rho \left( q,\,p\right) \geq 0\,,\qquad \int \rho \left( q,\,p\right)
\,d^{n}p=P(q)\,,\qquad \int \rho \left( q,\,p\right) \,d^{n}q=\widetilde{P}%
(p)\, 
\]
$P(q)$ and $\widetilde{P}(p)$ being the probability distributions for
position and momentum (the{\it \ marginals }of{\it \ }$\rho $).

The function $\rho \left( q,\,p\right) $ is normalized 
\begin{equation}
\int \rho \left( q,\,p\right) \,d^{n}q\,d^{n}p=1\,  \label{csm1}
\end{equation}
We consider an observable $X\left( q,\,p\right) $, that is, a function on
the phase space of the system. The inverse Fourier transform of the
characteristic function $\langle e^{ik\bullet X}\rangle $ for any vector
observable $X\left( q,\,p\right) $%
\begin{equation}
w\left( Y\right) =\frac{1}{\left( 2\,\pi \right) ^{n}}\,\int \langle
e^{ik\bullet X}\rangle e^{-ik\bullet Y}\,d^{n}k  \label{csm4}
\end{equation}
is a real nonnegative function which is normalized, since 
\begin{equation}
w\left( Y\right) =\int \rho \left( q,\,p\right) \,\delta ^{n}\left(
X(q,\,p)-Y\right) \,d^{n}qd^{n}p\,  \label{csm6}
\end{equation}
and 
\begin{equation}
\int w\left( Y\right) \,d^{n}Y=\int \rho \left( q,\,p\right)
\,d^{n}q\,d^{n}p=1\,  \label{csm9}
\end{equation}
As a classical analog of the quantum symplectic tomography observable,
introduced in \cite{Mancini95}, we consider the following classical
observable\cite{OlgaJRLR97} 
\begin{equation}
X\left( q,\,p\right) =\mu \circledast q+\nu \circledast p\,  \label{csm10}
\end{equation}
where $\circledast $ denotes the componentwise product of vectors 
\[
\left( \mu \circledast q\right) _{i}=\mu _{i}q_{i} 
\]
and $\mu $ and $\nu $ are vector-valued real parameters. Together with 
\[
P\left( q,\,p\right) =\frac{1}{\nu }\circledast \mu \circledast q+\left( 
\frac{1}{\mu }+{\bf 1}\right) \circledast p 
\]
(\ref{csm10}) is a symplectic transformation\footnote{%
This is not, of course, the most general symplectic transformation. In
general $X^{i}(x,p)=\mu _{k}^{i}x^{k}+\nu _{k}^{i}p^{k}$ with the
corresponding expression for $P(x,p)$. Here we have considered the
particular case where the tensors $\mu $ and $\nu $ are diagonal. That is
the reason for the non-covariant look of our equations.} of the position and
momentum observables. (${\bf 1}$ stands for the unit vector in $R^{n}$)

The vector variable $X\left( q,\,p\right) $ may be interpreted as a
coordinate of the system, when measured in a rotated and scaled reference
frame in the classical phase space. For the coordinate (\ref{csm10}) in the
transformed reference frame, we obtain from Eq.(\ref{csm4}) the distribution
function (the tomography map) 
\begin{equation}
w\left( X,\,\mu ,\,\nu \right) =\frac{1}{\left( 2\,\pi \right) ^{n}}\int
e^{-ik\bullet (X-\mu \circledast q-\nu \circledast p)}\,\rho \left(
q,\,p\right) \,d^{n}q\,d^{n}p\,d^{n}k  \label{csm11}
\end{equation}
This function is homogeneous 
\begin{equation}
w\left( \lambda X,\,\lambda \mu ,\,\lambda \nu \right) =|\lambda
|^{-n}w\left( X,\,\mu ,\,\nu \right)  \label{new1a}
\end{equation}
and Eq.(\ref{csm11}) has an inverse 
\begin{equation}
\rho \left( q,\,p\right) =\frac{1}{\left( 4\,\pi ^{2}\right) ^{n}}\int
w\left( X,\,\mu ,\,\nu \right) \,\exp \left[ i\left( X-\mu \circledast q-\nu
\circledast p\right) \bullet {\bf 1}\right] \,d^{n}X\,d^{n}\mu \,d^{n}\nu \,.
\label{csm22}
\end{equation}

Since the map 
\[
\rho \left( q,\,p\right) \Longrightarrow w\left( X,\,\mu ,\,\nu \right) 
\]
is invertible, the information contained in the distribution function $\rho
\left( q,\,p\right) $ is equivalent to the information contained in the
marginal distributions $w\left( X,\,\mu ,\,\nu \right) $.

The Boltzman evolution equation for the classical distribution function for
a particle with mass $m=1$ and potential $V(q)$, 
\begin{equation}
\frac{\partial \rho \left( q,\,p,\,t\right) }{\partial t}+p\bullet \nabla
_{q}\rho \left( q,\,p,\,t\right) -\nabla _{q}V(q)\bullet \nabla _{p}\rho
\left( q,\,p,\,t\right) \,=0  \label{csm23}
\end{equation}
can be rewritten in terms of the marginal distribution $w\left( X,\,\mu
,\,\nu ,\,t\right) $%
\begin{equation}
\frac{\partial w}{\partial t}-\mu \bullet \nabla _{\nu }w-\nabla _{x}V\left(
-\nabla _{X}^{-1}\circledast \nabla _{\mu }\right) \bullet \left( \nu
\circledast \nabla _{X}w\right) =0\,  \label{csm24}
\end{equation}

For the mean value of the position and momentum we have 
\begin{eqnarray}
\left\langle \left( 
\begin{array}{l}
q \\ 
p
\end{array}
\right) \right\rangle &=&\int \rho \left( q,\,p\right) \left( 
\begin{array}{l}
q \\ 
p
\end{array}
\right) \,d^{n}q\,d^{n}p  \label{1.11} \\
&=&i\,\int w\left( X,\,\mu ,\,\nu \right) e^{iX\bullet {\bf 1}}\left( 
\begin{array}{l}
\nabla _{\mu } \\ 
\nabla _{\nu }
\end{array}
\right) \left( \delta ^{n}(\mu )\,\delta ^{n}\left( \nu \right) \right)
\,d^{n}X\,d^{n}\mu \,d^{n}\nu  \nonumber
\end{eqnarray}
With a change of variables $X\rightarrow \mu X$ and the homogeneity property 
$w(\mu X,\mu ,0)=\mu ^{-n}w\left( X,{\bf 1},0\right) $ one may, for example,
check the consistency of this definition for the mean value of the position 
\begin{equation}
\langle q\rangle =\int w\left( X,{\bf 1},0\right) X\,d^{n}X\,  \label{new1b}
\end{equation}

By $\Pi _{{\rm cl}}\left( X,\,\mu ,\,\nu ,\,X^{\prime },\,\mu ^{\prime
},\,\nu ^{\prime },\,t_{2},\,t_{1}\right) $ we denote the classical
propagator that connects two marginal distributions at different times $%
t_{0} $ and $t$ $\left( t>t_{0}\right) $%
\begin{equation}
w\left( X,\mu ,\nu ,t\right) =\int \Pi _{{\rm cl}}\left( X,\mu ,\nu
,X^{\prime },\mu ^{\prime },\nu ^{\prime },t,t_{0}\right) w\left( X^{\prime
},\mu ^{\prime },\nu ^{\prime },t_{0}\right) \,d^{n}X^{\prime }\,d^{n}\mu
^{\prime }\,d^{n}\nu ^{\prime }.  \label{1.12}
\end{equation}
The propagator satisfies the equation 
\begin{equation}
\frac{\partial \Pi _{{\rm cl}}}{\partial t_{2}}-\mu \bullet \nabla _{\nu
}\Pi _{{\rm cl}}-\nabla _{x}V\left( -\nabla _{X}^{-1}\circledast \nabla
_{\mu }\right) \bullet \left( \nu \circledast \nabla _{X}\Pi _{{\rm cl}%
}\right) =0  \label{1.13}
\end{equation}
with boundary condition 
\begin{equation}
\lim_{t_{2}\rightarrow t_{1}}\Pi _{{\rm cl}}\left( X,\mu ,\nu ,X^{\prime
},\mu ^{\prime },\nu ^{\prime },t_{2},t_{1}\right) =\delta ^{n}\left(
X-X^{\prime }\right) \delta ^{n}\left( \mu -\mu ^{\prime }\right) \delta
^{n}\left( \nu -\nu ^{\prime }\right)  \label{1.14}
\end{equation}

\subsection{Quantum mechanics}

For quantum mechanics the construction is similar and the mathematical
nature of the quantities that are constructed is the same, because it is a
general fact that the inverse Fourier transform of a characteristic function
is a positive distribution. The marginal distributions that are obtained are
simply related to other well known quantum mechanical quantities. It was
shown\cite{Mancini95} that for the generic linear combination 
\begin{equation}
X=\mu \circledast q+\nu \circledast p  \label{X}
\end{equation}
where $q$ and $p$ are the position and the momentum, the marginal
distribution $w\,(X,\,\mu ,\,\nu )$ (normalized in the variable $X$ and
depending on two vector-valued real parameters $\mu $ and $\nu $) is related
to the Wigner function $W(q,\,p)$. For $n$ degrees of freedom one has 
\begin{equation}
w\left( X,\,\mu ,\,\nu \right) =\int \exp \left[ -ik\bullet (X-\mu
\circledast q-\nu \circledast p)\right] W(q,\,p)\,\frac{d^{n}k\,d^{n}q%
\,d^{n}p}{(4\pi ^{2})^{n}}\,  \label{w}
\end{equation}
We see that Eq.(\ref{w}) is formally identical to (\ref{csm11}) of the
classical case. For a pure state with wave function $\Psi \left( y\right) $,
the marginal distribution would be\cite{MendesManko} 
\begin{equation}
w\left( X,\,\mu ,\nu \right) =\frac{1}{\left( 2\,\pi \right) ^{n}|\nu
_{1}\cdots \nu _{n}|}\left| \int \Psi \left( y\right) \exp
i\sum_{j=1}^{n}\left( \frac{\mu _{j}y_{j}^{2}}{2\nu _{j}}-\frac{y_{j}X_{j}}{%
\nu _{j}}\right) \,d^{n}y\right| ^{2}  \label{wp}
\end{equation}
Eq.(\ref{w}) may be inverted and the Wigner function expressed in terms of
the marginal distribution, like in the classical case of Eq.(\ref{csm22}) 
\begin{equation}
W(q,\,p)=\left( \frac{1}{2\pi }\right) ^{n}\int w\left( X,\,\mu ,\,\nu
\right) \exp \left[ i\left( X-\mu \circledast q-\nu \circledast p\right)
\bullet {\bf 1}\right] \,d^{n}\mu \,d^{n}\nu \,d^{n}X\,  \label{W}
\end{equation}
Therefore the usual quantum mechanical quantities can be reconstructed from
the marginal distributions. These quantities (wave function and Wigner
function) have a nature quite different from the classical quantities,
however the marginals $w\left( X,\,\mu ,\,\nu \right) $ are in both cases
positive distributions with the same physical meaning.

As was shown in~\cite{ManciniPL}, for a system with Hamiltonian 
\begin{equation}
H=\frac{{p}^{2}}{2}+V(q)\,,  \label{HV}
\end{equation}
the marginal distribution satisfies the quantum time-evolution equation 
\begin{equation}
\begin{array}{r}
\frac{\partial w}{\partial t}-\mu \bullet \nabla _{\nu }w-i\frac{1}{\hbar }%
\{V\left( -\nabla _{X}^{-1}\circledast \nabla _{\mu }-i\frac{1}{2}\hbar \nu
\circledast \nabla _{X}\right) \\ 
-V\left( -\nabla _{X}^{-1}\circledast \nabla _{\mu }+i\frac{1}{2}\hbar \nu
\circledast \nabla _{X}\right) \}w=0
\end{array}
\label{FP}
\end{equation}
which provides a dynamical characterization of quantum dynamics, alternative
to the Schr\"{o}dinger equation.

The evolution equation~(\ref{FP}) can also be written in the form 
\begin{equation}
\begin{array}{l}
\frac{\partial w}{\partial t}-\mu \bullet \nabla _{\nu }w-\nabla _{x}V\left( 
\widetilde{q}\right) \bullet \left( \nu \circledast \nabla _{X}w\right) \\ 
+\frac{2}{\hbar }\sum_{n=1}^{\infty }(-1)^{n+1}\left( \frac{\hbar }{2}%
\right) ^{2n+1}\frac{\nabla _{i_{1}\cdots i_{2n+1}}V\left( \widetilde{q}%
\right) }{(2n+1)!}\left( \nu \circledast \nabla _{X}\right) _{i_{1}}\cdots
\left( \nu \circledast \nabla _{X}\right) _{i_{2n+1}}w=0
\end{array}
\label{defeq}
\end{equation}
where $\widetilde{q}$ stands for the operator 
\[
\widetilde{q}=-\nabla _{X}^{-1}\circledast \nabla _{\mu } 
\]
and a sum over repeated indices is implied.

In Moyal's\cite{Moyal49} formulation of quantum mechanics in phase-space,
the transition from the classical to the quantum structure is a deformation
of the Poisson algebra\cite{Lichne1} with deformation parameter $\hbar $. In
the symplectic tomography formulation, that we are describing, classical and
quantum mechanics are described by the same set of positive probability
distributions $w\left( X,\,\mu ,\,\nu \right) $, the $\hbar -$deformation
appearing only in the time-evolution equation (\ref{defeq}).

For the propagator 
\begin{equation}
w\left( X,\mu ,\nu ,t\right) =\int \Pi \left( X,\mu ,\nu ,X^{\prime },\mu
^{\prime },\nu ^{\prime },t,t_{0}\right) w\left( X^{\prime },\mu ^{\prime
},\nu ^{\prime },t_{0}\right) \,d^{n}X^{\prime }\,d^{n}\mu ^{\prime
}\,d^{n}\nu ^{\prime }.  \label{2.7}
\end{equation}
the equation is 
\begin{equation}
\begin{array}{l}
\frac{\partial \Pi }{\partial t}-\mu \bullet \nabla _{\nu }\Pi -\nabla
_{x}V\left( \widetilde{q}\right) \bullet \left( \nu \circledast \nabla
_{X}\Pi \right) \\ 
+\frac{2}{\hbar }\sum_{n=1}^{\infty }(-1)^{n+1}\left( \frac{\hbar }{2}%
\right) ^{2n+1}\frac{\nabla _{i_{1}\cdots i_{2n+1}}V\left( \widetilde{q}%
\right) }{(2n+1)!}\left( \nu \circledast \nabla _{X}\right) _{i_{1}}\cdots
\left( \nu \circledast \nabla _{X}\right) _{i_{2n+1}}\Pi \\ 
=0
\end{array}
\label{2.8}
\end{equation}
with boundary condition 
\begin{equation}
\lim_{t\rightarrow t_{0}}\Pi \left( X,\mu ,\nu ,X^{\prime },\mu ^{\prime
},\nu ^{\prime },t,t_{0}\right) =\delta ^{n}\left( X-X^{\prime }\right)
\delta ^{n}\left( \mu -\mu ^{\prime }\right) \delta ^{n}\left( \nu -\nu
^{\prime }\right)  \label{2.9}
\end{equation}

\section{Lyapunov exponents}

\subsection{Density formulation in classical mechanics}

Lyapunov exponents and other ergodic invariants in the classical theory are
usually formulated in terms of quantities related to trajectories in
phase-space, like tangent maps, refinement of partitions, etc.\cite{Mane}.
Here, as a preparation for the formulation of Lyapunov exponents in quantum
mechanics, using the marginal distributions $w\left( X,\,\mu ,\,\nu \right) $%
, we explain briefly how these quantities may, in classical mechanics, be
expressed as functionals of phase-space densities rather than in terms of
trajectories. For more details we refer to \cite{Vilela2}.

A density in phase-space is a non-negative, normalized, integrable function,
the space of densities being denoted by $D$%
\begin{equation}
D=\left\{ \rho \in L^{1}:\rho \geq 0,\left\| \rho \right\| _{1}=1\right\}
\label{3.1}
\end{equation}
$D$ is the space of functions that, by the Radon-Nikodym theorem,
characterize the measures that are absolutely continuous with respect to the
underlying measure in phase-space. However, to define Lyapunov exponents by
densities, it is necessary to restrict oneself to a subspace of {\it %
admissible densities} defined as follows:

To each $\rho \in D$ we associate a square root, that becomes an element of
an $L^{2}$ space. We then construct a Gelfand triplet 
\begin{equation}
E^{*}\supset L^{2}\supset E  \label{3.2}
\end{equation}
where $E$ is the space of functions of rapid decrease topologized by the
family of semi-norms $\left\| x_{\alpha }\partial _{\beta }f\right\| _{2}$
and $E^{*}$ is its dual. Because $E$ is an algebra $f\in E$ implies $%
f^{2}\in E$. Therefore for each $f$ such that $\left\| f\right\| _{2}=1$ , $%
\rho =f^{2}$ is an admissible density. The restriction to such a subspace of
admissible densities is necessary to be able to define Gateaux derivatives
along generalized functions with point support. Gateaux derivatives along
derivatives of the delta function play for densities the same role as the
tangent map for trajectories. In this setting the Lyapunov exponent is\cite
{Vilela2} 
\begin{equation}
\lambda _{v}=\lim_{t\rightarrow \infty }\frac{1}{t}\log \left\|
-v^{i}D_{\partial _{i}\delta _{x}}\left( \int d\mu (y)yP^{t}\rho (y)\right)
\right\|  \label{3.3}
\end{equation}
$v\in R^{2n}$, $\left\| \cdot \right\| $ is the vector norm and the Gateaux
derivative $D_{\partial _{i}\delta _{x}}$ operates in the argument of the
functional, that is, on the initial density 
\begin{equation}
D_{\partial _{i}\delta _{x}}F\left( \rho (y)\right) =\lim_{\varepsilon
\rightarrow 0}\frac{1}{\varepsilon }\left\{ F\left( \rho (y)+\varepsilon
\partial _{i}\delta (y-x)\right) -F\left( \rho (y)\right) \right\}
\label{3.4}
\end{equation}
$\mu $ is the invariant measure in the support of which the Lyapunov
exponent is being defined and $P^{t}$ is the operator of time evolution for
densities 
\begin{equation}
P^{t}\rho (y,0)=\rho (y,t)  \label{3.5}
\end{equation}

A simple computation shows that the expression (\ref{3.3}) is equivalent to
the usual definition of Lyapunov exponent in terms of trajectories and the
tangent map 
\begin{equation}
\lambda _{v}=\lim_{t\rightarrow \infty }\frac{1}{t}\log \left\|
DT_{x}^{t}v\right\|  \label{3.6}
\end{equation}
where $DT_{x}^{t}$ stands for the evolved tangent map applied to the vector $%
v$ at the phase-space point $x$. Here and in Eq.(\ref{3.3}) $x$ and $y$ are
phase space vectors, that is, in the notation of Sect. 2, $x=\left(
q_{x},p_{x}\right) $ and $y=\left( q_{y},p_{y}\right) $.

According to Oseledec theorem \cite{Oseledec} \cite{Raghu}, for $\mu -$%
almost every point $x$ there is a decreasing sequence of vector spaces 
\[
R^{2n}=E_{1}(x)\supset E_{2}(x)\supset \cdots \supset E_{r}=\{0\} 
\]
such that, by choosing the vector $v$ in $E_{s}(x)\setminus E_{s+1}(x)$, the 
$s$th Lyapunov exponent is obtained by the above calculation.

A similar construction is possible for the metric entropy. For the entropy,
the notion that seems more appropriate for generalization to quantum
mechanics\cite{Vilela3} \cite{Vilela2} is the Brin-Katok local entropy\cite
{Mane}, which for the classical case and for a compact metric space is
equivalent to the Kolmogoroff-Sinai entropy. It is defined as follows:

Define 
\[
B_{\varepsilon }\left( T,t,x\right) =\left\{ y:d\left( T^{\tau }(x),T^{\tau
}(y)\right) \leq \varepsilon {\ ,}0\leq \tau \leq t\right\} 
\]
$B_{\varepsilon }\left( T,t,x\right) $ is the ball of phase-space points
around $x$ that, in the course of time evolution, do not separate by a
distance larger than $\varepsilon $ up to time $t$. $T^{\tau }(x)$ is the
image of $x$ after the time $\tau $ and $d(\cdot ,\cdot )$ is the distance.
The local entropy $h(T,x)$ measures the weighed (in the $\mu $-measure) rate
of shrinkage in time of the ball $B_{\varepsilon }\left( T,t,x\right) $,
namely 
\[
h(T,x)=\lim_{\varepsilon \rightarrow 0}\lim_{t\rightarrow \infty }\left\{ -%
\frac{1}{t}\log \mu \left( B_{\varepsilon }\left( T,t,x\right) \right)
\right\} 
\]
As in the case of the Lyapunov exponent, this quantity may be expressed as a
functional of (admissible) densities by rewriting the ball $B_{\varepsilon
}\left( T,t,x\right) $ as 
\[
B_{\varepsilon }\left( T,t,x\right) =\left\{ y:\left| D_{(\delta _{x}-\delta
_{y})}\left( \int d\mu (z)zP^{\tau }\rho (z)\right) \right| \leq \varepsilon 
{\ };0\leq \tau \leq t\right\} 
\]

\subsection{Classical and quantum Lyapunov exponents by marginal
distributions}

Let us now translate the equations of the preceding subsection in the
tomographic framework discussed in Section~2. Initial densities are, by the
tomographic map, mapped to initial tomographic densities by (\ref{csm11}) 
\begin{equation}
\rho \left( q,p\right) \rightarrow w\left( X,\mu ,\nu ,t=0\right) \equiv
w\left( X,\mu ,\nu \right)  \label{L8}
\end{equation}
To compute the Gateaux derivatives notice that the generalized density ($\in
E^{*}$) 
\begin{equation}
\left( v_{1}\bullet \nabla _{q}+v_{2}\bullet \nabla _{p}\right) \left\{
\delta ^{n}\left( q-q_{0}\right) \,\delta ^{n}(p-p_{0})\right\}  \label{4.1}
\end{equation}
is mapped to the tomographic generalized density $w_{\eta }$($\in E^{*}$) 
\begin{equation}
w_{\eta }\left( X,\mu ,\nu \right) =\left( \left( v_{1}\circledast \mu
+v_{2}\circledast \nu \right) \bullet \nabla _{X}\right) \delta ^{n}\left(
X-\mu q_{0}-\nu p_{0}\right)  \label{L9}
\end{equation}
According to Eq.(\ref{3.3}), to compute the Lyapunov exponent one has to
obtain the expectation value of a generic phase-space vector on the
time-evolved perturbation of the initial density (\ref{4.1}). Therefore 
\begin{equation}
\lambda _{v}=\lim_{t\rightarrow \infty }\frac{1}{t}\log \left\| 
\begin{array}{c}
\int d^{n}qd^{n}p\left( 
\begin{array}{l}
q \\ 
p
\end{array}
\right) {\cal K}\left( q,p,q^{\prime },p^{\prime },t\right) \left(
v_{1}\bullet \nabla _{q^{\prime }}+v_{2}\bullet \nabla _{p^{\prime }}\right)
\\ 
\delta ^{n}\left( q^{\prime }-q_{0}\right) \,\delta ^{n}(p^{\prime
}-p_{0})dq^{^{\prime }}dp^{^{\prime }}
\end{array}
\right\|  \label{4.2}
\end{equation}
where ${\cal K}\left( q,p,q^{\prime },p^{\prime },t\right) $ is the
evolution kernel for densities 
\[
\rho \left( q,p,t\right) =\int {\cal K}\left( q,p,q^{\prime },p^{\prime
},t\right) \rho \left( q^{\prime },p^{\prime }\right) \,d^{n}q^{\prime
}\,d^{n}p^{\prime } 
\]
Notice that in Eq.(\ref{4.2}) the integration is carried over the flat
phase-space measure $d^{n}qd^{n}p$. The result is equivalent to (\ref{3.3})
for an invariant measure absolutely continuous with respect to $d^{n}qd^{n}p$%
. However the information and the dependence of the Lyapunov exponent on the
invariant measure is carried by the choice of the initial point $\left(
q_{0},p_{0}\right) $. The set of Lyapunov exponents that is obtained by (\ref
{4.2}) is therefore the one that corresponds to the invariant measure on
whose support $\left( q_{0},p_{0}\right) $ lies.

Using (\ref{1.11}), Eq.(\ref{4.2}) may now be rewritten using marginal
distributions 
\begin{equation}
\lambda _{v}=\lim_{t\rightarrow \infty }\frac{1}{t}\log \left\| 
\begin{array}{c}
\int d^{n}X\,d^{n}\mu \,d^{n}\nu e^{iX\bullet {\bf 1}}\left( \left( 
\begin{array}{l}
\nabla _{\mu } \\ 
\nabla _{\nu }
\end{array}
\right) \delta ^{n}(\mu )\delta ^{n}\left( \nu \right) \right) \Pi _{{\rm cl}%
}\left( X,\,\mu ,\,\nu ,\,X^{\prime },\,\mu ^{\prime },\,\nu ^{\prime
},\,t,\,0\right) \\ 
\left( \left( v_{1}\circledast \mu ^{\prime }+v_{2}\circledast \nu ^{\prime
}\right) \bullet \nabla _{X^{\prime }}\right) \delta ^{n}\left( X^{\prime
}-\mu ^{\prime }q_{0}-\nu ^{\prime }p_{0}\right) dX^{\prime n}d\mu ^{\prime
n}d\nu ^{\prime n}
\end{array}
\right\|  \label{4.3}
\end{equation}
where $\Pi _{{\rm cl}}\left( X,\,\mu ,\,\nu ,\,X^{\prime },\,\mu ^{\prime
},\,\nu ^{\prime },\,t_{2},\,t_{1}\right) $ is the classical propagator
defined in (\ref{1.12}) - (\ref{1.14}).

Because (\ref{3.3}) is equivalent to the usual definition of Lyapunov
exponent, Eq.(\ref{4.3}), being equivalent to (\ref{3.3}), is also a correct
expression for the classical Lyapunov exponent.

Now the transition to quantum mechanics is straightforward. Marginal
distributions in classical and quantum mechanics satisfy formally identical
expressions and have the same physical interpretation as probability
densities. The only difference lies on the time-evolution which in classical
mechanics obeys Eq.(\ref{csm24}) and in quantum mechanics the $\hbar $%
-deformed equation (\ref{FP}). Therefore the Lyapunov exponent in quantum
mechanics will also be given by equation (\ref{4.3}), with however the
classical propagator $\Pi _{{\rm cl}}$ replaced by the quantum propagator $%
\Pi $ for marginal distributions, defined in (\ref{2.7}) - (\ref{2.9}).

\section{Hilbert space expression for the quantum Lyapunov exponent}

As we will see in Sect.5, Eq.(\ref{4.3}) provides an efficient way to
compute the Lyapunov exponent. However, for comparison with other
approaches, it is useful to translate Eq.(\ref{4.3}) in the Hilbert space
quantum mechanical formalism.

To simplify the notation, we consider $n=1$, that is, a two-dimensional
phase space. Generalization to the $n-$dimensional case is straightforward.
To write the quantum Lyapunov exponent (\ref{4.3}) in terms of the evolution
operator acting in the Hilbert space of states, we use the following equation%
\cite{Olga2} that relates the tomographic propagator to Hilbert space
Green's functions 
\begin{eqnarray*}
\Pi \left( X,\mu ,\nu ,X^{\prime },\mu ^{\prime },\nu ^{\prime },t\right) &=&%
\frac{1}{4\,\pi ^{2}}\int k^{2}\,G\left( a+\frac{k\nu }{2},y,t\right)
\,G^{*}\left( a-\frac{k\nu }{2},z,t\right) \\
&&\times \delta \left( y-z-k\nu ^{\prime }\right) \,\exp \left[ ik\left(
X^{\prime }-X+\mu a-\mu ^{\prime }\,\frac{y+z}{2}\right) \right] \\
&&\,dk\,dy\,dz\,da
\end{eqnarray*}

Then, the quantum Lyapunov exponent expressed in terms of Green's functions
is 
\begin{eqnarray*}
\lambda _{v} &=&\lim_{t\to \infty }\frac{1}{t}\int
da~dz~e^{-2ip_{0}z}G^{*}(a,z,t)v_{1}\left[ ip_{0}G(a,2q_{0}-z,t)-\frac{%
\partial G}{\partial z}(a,2q_{0}-z,t)\right] \\
&&+iv_{2}(q_{0}-z)G(a,2q_{0}-z,t)
\end{eqnarray*}
Using a complete set of wave functions $\psi _{n}(x,t)$ this expression may
be rewritten 
\begin{eqnarray*}
\lambda _{v} &=&\lim_{t\to \infty }\frac{1}{t}\ln |\int
da~dz~e^{-2ip_{0}z}a\sum_{n,m}\psi _{n}^{*}(a,t)\psi _{n}(z,t) \\
&&\{i[v_{1}p_{0}+v_{2}(q_{0}-z)][\psi _{m}(a,t)\psi _{m}^{*}(2q_{0}-z,t)] \\
&&+v_{1}\psi _{m}(a,t)\psi _{m}^{*\prime }(2q_{0}-z,t)\}|
\end{eqnarray*}

In Refs.\cite{Mendes1}-\cite{Vilela4} a quantum characteristic exponent has
been defined, in Hilbert space, by using the time evolution of the
expectation values on a $\delta ^{^{\prime }}-$perturbed wave function. This
definition may be compared with the present marginal probability
construction. Under a $\varepsilon \delta ^{^{\prime }}\left( x-q_{0}\right) 
$ perturbation of the wave function, the density matrix changes, in leading
order, by 
\begin{equation}
\Delta \left( \psi (x)\psi (x^{^{\prime }})\right) =\varepsilon \left\{
\delta ^{^{\prime }}\left( x-q_{0}\right) \psi ^{*}(x^{^{\prime }})+\psi
(x)\delta ^{^{\prime }}\left( x^{^{\prime }}-q_{0}\right) \right\}
\label{5.6}
\end{equation}
On the other hand the marginal probability perturbation 
\[
\left( v_{1}\mu +v_{2}\nu \right) \delta ^{^{\prime }}\left( X-\mu q_{0}-\nu
p_{0}\right) 
\]
induces a perturbation of the density matrix 
\begin{equation}
\Delta \rho \left( x,x^{^{\prime }}\right) =-e^{ip_{0}\left( x-x^{^{\prime
}}\right) }\left\{ v_{1}\delta ^{^{\prime }}\left( q_{0}-\frac{x+x^{^{\prime
}}}{2}\right) +iv_{2}\left( x-x^{^{\prime }}\right) \delta \left( q_{0}-%
\frac{x+x^{^{\prime }}}{2}\right) \right\}  \label{5.8}
\end{equation}
Comparing (\ref{5.6}) and (\ref{5.8}) one sees that, for the calculation of
the Lyapunov exponent, they coincide on the diagonal terms $\left(
x=x^{^{\prime }}\right) $, but are different on the non-diagonal terms. In
particular, the marginal density perturbation is not a pure state
perturbation and cannot be reproduced by the perturbation of a single wave
function.

\section{Examples: Kicked systems on the line and on the circle}

\subsection{One-dimensional systems with time-dependent potentials}

We consider here one-dimensional systems with time-dependent potentials
defined by the Hamiltonian 
\begin{equation}
H=\frac{p^{2}}{2}+V(q,t)  \label{6.1}
\end{equation}
For these systems, the Lyapunov exponent expression (\ref{4.3}) is 
\begin{equation}
\lambda _{v}=\lim_{t\rightarrow \infty }\frac{1}{t}\log \left\| \int
dX\,d\mu \,d\nu e^{iX}\left( \left( 
\begin{array}{l}
\partial _{\mu } \\ 
\partial _{\nu }
\end{array}
\right) \delta (\mu )\delta \left( \nu \right) \right) F\left( X,\mu ,\nu
,t\right) \right\|  \label{6.2}
\end{equation}
where $F\left( X,\mu ,\nu ,t\right) $ is the time-evolved perturbation,
namely 
\begin{equation}
F\left( X,\mu ,\nu ,t\right) =\int \Pi \left( X,\,\mu ,\,\nu ,\,X^{\prime
},\,\mu ^{\prime },\,\nu ^{\prime },\,t,\,0\right) \left( v_{1}\mu ^{\prime
}+v_{2}\nu ^{\prime }\right) \delta ^{^{\prime }}\left( X^{\prime }-\mu
^{\prime }q_{0}-\nu ^{\prime }p_{0}\right) dX^{^{\prime }}d\mu ^{^{\prime
}}d\nu ^{\prime }  \label{6.3}
\end{equation}
Passing to the Fourier transform 
\begin{equation}
G\left( k,\mu ,\nu ,t\right) =\frac{1}{2\pi }\int e^{ikX}F\left( X,\mu ,\nu
,t\right) dX  \label{6.4}
\end{equation}
one obtains 
\begin{eqnarray}
\lambda _{v} &=&\lim_{t\rightarrow \infty }\frac{1}{t}\log \left\| \int d\mu
\,d\nu \left( \left( 
\begin{array}{l}
\partial _{\mu } \\ 
\partial _{\nu }
\end{array}
\right) \delta (\mu )\delta \left( \nu \right) \right) G\left( 1,\mu ,\nu
,t\right) \right\|  \label{6.5} \\
&=&\lim_{t\rightarrow \infty }\frac{1}{t}\log \left\| \left( 
\begin{array}{l}
G^{(2)}\left( 1,0,0,t\right) \\ 
G^{(3)}\left( 1,0,0,t\right)
\end{array}
\right) \right\|  \nonumber
\end{eqnarray}
where by $G^{(2)}$ and $G^{(3)}$ we denote the derivatives in the second and
third arguments and $G\left( k,\mu ,\nu ,t\right) $ is a solution of the
equation 
\begin{equation}
\begin{array}{l}
\frac{\partial G}{\partial t}-\mu \frac{\partial G}{\partial \nu }-ik\nu
\partial _{q}V\left( -\frac{1}{ik}\frac{\partial }{\partial \mu }\right) G
\\ 
+\frac{2}{\hbar }\sum_{n=1}^{\infty }\frac{(-1)^{n+1}}{(2n+1)!}\left( ik%
\frac{\nu \hbar }{2}\right) ^{2n+1}\frac{\partial ^{2n+1}}{\partial q^{2n+1}}%
V\left( -\frac{1}{ik}\frac{\partial }{\partial \mu }\right) G=0
\end{array}
\label{6.6}
\end{equation}
with initial condition 
\begin{equation}
G\left( k,\mu ,\nu ,t\right) =-\frac{ik}{2\pi }\left( v_{1}\mu +v_{2}\nu
\right) e^{ik\left( q_{0}\mu +p_{0}\nu \right) }  \label{6.7}
\end{equation}
Therefore, the computation of the Lyapunov exponents, both classical and
quantum, reduces to the study of the large time limit of the solutions of
Eq.(\ref{6.6}). Also the simple expression (\ref{6.5}) shows that, despite
its apparently complex form, Eq.(\ref{4.3}) is a computationally efficient
way to obtain the Lyapunov exponent.

We now study several time-dependent (kicked) potentials.

\subsection{Harmonic kicks on the line}

Here the potential is

\begin{equation}
V\left( q\right) =\frac{\gamma \alpha }{\pi }\frac{q^{2}}{2}\sum_{n=-\infty
}^{\infty }\delta \left( t-n\right)  \label{6.8}
\end{equation}
This system belongs to the class of time-dependent quadratic systems\cite
{OlgaJRLR97} and a solution may be found for the general case 
\begin{equation}
V\left( q\right) =\alpha \left( t\right) \frac{q^{2}}{2}  \label{6.9}
\end{equation}
Eq.(\ref{6.6}) reduces to 
\begin{equation}
\frac{\partial G}{\partial t}-\mu \frac{\partial G}{\partial \nu }+\alpha
\left( t\right) \nu \frac{\partial G}{\partial \mu }=0  \label{6.10}
\end{equation}
The $\hbar -$deformed part of Eq.(\ref{6.6}) disappears and we obtain the
(expected) result that, in this case, classical and quantum results
coincide. Eq.(\ref{6.10}) has the solution 
\begin{equation}
G\left( k,\mu ,\nu ,t\right) =G\left( k,\frac{\mu }{2}\left( \varepsilon
+\varepsilon ^{*}\right) +\frac{\nu }{2}\left( \stackrel{\bullet }{%
\varepsilon }+\stackrel{\bullet }{\varepsilon }^{*}\right) ,\frac{\nu }{2i}%
\left( \stackrel{\bullet }{\varepsilon }-\stackrel{\bullet }{\varepsilon }%
^{*}\right) +\frac{\mu }{2i}\left( \varepsilon -\varepsilon ^{*}\right)
,0\right)  \label{6.11}
\end{equation}
the function $\varepsilon \left( t\right) $ being a solution of 
\begin{equation}
\stackrel{\bullet \bullet }{\varepsilon }\left( t\right) +\alpha \left(
t\right) \varepsilon \left( t\right) =0  \label{6.12}
\end{equation}
with initial conditions 
\begin{equation}
\varepsilon \left( 0\right) =1,\quad \stackrel{\bullet }{\varepsilon }\left(
0\right) =i  \label{6.13}
\end{equation}
For the calculation of the Lyapunov exponent, the function $G\left( k,\cdot
,\cdot ,0\right) $ is the one given in Eq.(\ref{6.7}).

For the kicked case in (\ref{6.8}) the function $\varepsilon \left( t\right) 
$ is obtained by establishing a matrix recurrence relation. Between times $%
t_{n-1}$ and $t_{n}$ we denote the function $\varepsilon \left( t\right) $
by $\varepsilon _{n}\left( t\right) $. Then 
\begin{equation}
\varepsilon _{n}(t)=a_{n}+b_{n}t\,.  \label{6.14}
\end{equation}
and the following matrix recurrence is obtained 
\begin{equation}
\left( 
\begin{array}{c}
a_{n} \\ 
b_{n}
\end{array}
\right) =\left( 
\begin{array}{cl}
1+\frac{\gamma \alpha }{\pi }n & \frac{\gamma \alpha }{\pi }n^{2} \\ 
-\frac{\gamma \alpha }{\pi } & 1-\frac{\gamma \alpha }{\pi }n
\end{array}
\right) \left( 
\begin{array}{c}
a_{n-1} \\ 
b_{n-1}
\end{array}
\right)  \label{6.15}
\end{equation}
with initial condition 
\begin{equation}
a_{0}=1\,,\qquad b_{0}=i  \label{6.16}
\end{equation}
The recurrence relation (\ref{6.15}) for the coefficients $a_{n},b_{n}$
yields the map 
\begin{equation}
\left( 
\begin{array}{c}
{\cal E}(t_{n+1}) \\ 
\dot{{\cal E}}(t_{n+1})
\end{array}
\right) =\left( 
\begin{array}{c}
q_{n+1} \\ 
p_{n+1}
\end{array}
\right) =\left( 
\begin{array}{cl}
1 & 1 \\ 
-\frac{\gamma \alpha }{\pi } & 1-\frac{\gamma \alpha }{\pi }
\end{array}
\right) \left( 
\begin{array}{c}
q_{n} \\ 
p_{n}
\end{array}
\right)  \label{6.17}
\end{equation}
for the position and momentum just after each kick.

The Floquet matrix in (\ref{6.17}) has two eigenvalues 
\begin{equation}
\lambda _{1,2}=1-\frac{\gamma \alpha }{2\pi }\pm \sqrt{\frac{(\gamma \alpha
)^{2}}{4\pi ^{2}}-\frac{\gamma \alpha }{\pi }}\,.  \label{6.18}
\end{equation}
Substituting in Eq.(\ref{6.5}) one concludes that for $z=\gamma \alpha /\pi
<4$ the Lyapunov exponent vanishes and that for $z=\gamma \alpha /\pi >4$
there is one positive Lyapunov exponent 
\begin{equation}
\lambda =\ln \left| 1-\frac{1}{2}z-\sqrt{\frac{1}{4}z^{2}-z}\right| \,,
\label{6.19}
\end{equation}
This is just the classical result and also the quantum result for another
definition of quantum exponent\cite{Mendes1} \cite{Vilela4}, already
discussed in Section 4. The positive Lyapunov exponent corresponds simply to
the situation where the Floquet operator spectrum is transient absolutely
continuous\cite{Slawny}. However, the next example corresponds to a
classically chaotic system which, when quantized, yields an absolutely
continuous quasi-energy spectrum \cite{Weigert}. This suggests that it might
be an example of genuine quantum chaos. The Lyapunov exponent analysis
supports this conclusion.

\subsection{The configurational quantum cat}

The configurational quantum cat is a system with 4-dimensional phase space,
for which the configuration space dynamics resembles the classical Arnold cat%
\cite{Arnold}. It describes a charged particle constrained to move in the
unit square with periodic boundary conditions, under the influence of
time-dependent electromagnetic pulses. It may be associated to the
Hamiltonians\cite{Weigert} \cite{Mendes1} 
\begin{equation}
H_{1}=\frac{1}{2}p_{1}^{2}+\frac{1}{2}p_{2}^{2}+x_{2}p_{1}+x_{1}p_{2}\sum_{n%
\in Z}\delta \left( t-n\tau \right)  \label{qc1}
\end{equation}
or 
\begin{equation}
H_{1}=\frac{1}{2}p_{1}^{2}+\frac{1}{2}p_{2}^{2}+\left( x_{2}p_{1}+\left(
x_{1}+x_{2}\right) p_{2}\right) \sum_{n\in Z}\delta \left( t-n\tau \right)
\label{qc2}
\end{equation}
A similar model may be constructed\cite{Mendes1} by considering only the
kick part and defining the quantum theory directly by the Floquet operator.
The deal with the Hamiltonians (\ref{qc1}) or (\ref{qc2}) by the tomographic
formalism, we need to extend it to nonlocal potentials.

Let a nonlocal Hamiltonian be written as 
\begin{equation}
H(x,p)=T(p)+V(x)+I(x,p)  \label{1}
\end{equation}
where $T(p)$ is the kinetic energy, $V(x)$ the local potential energy and $%
I(x,p)$ the symmetrized position and momentum-dependent interaction. Then,
the equation for the density operator $\rho $ 
\begin{equation}
\dot{\rho}+i(H\rho -\rho H)=0  \label{2}
\end{equation}
becomes, in the tomographic representation 
\begin{equation}
\dot{w}({\bf X},\vec{\mu},\vec{\nu},t)+iH(\hat{{\bf x}}^{(1)},\hat{{\bf p}}%
^{(1)})w({\bf X},\vec{\mu},\vec{\nu},t)-iH(\hat{{\bf x}}^{(2)},-\hat{{\bf p}}%
^{(2)})w({\bf X},\vec{\mu},\vec{\nu},t)=0  \label{3}
\end{equation}
where, for $n$ degrees of freedom ${\bf X}=(X_{1},X_{2},\ldots ,X_{n})$, ${%
\vec{\mu}}=(\mu _{1},\mu _{2},\ldots ,\mu _{n})$, ${\vec{\nu}}=(\nu _{1},\nu
_{2},\ldots ,\nu _{n})$, and the components of the vector-operators $\hat{%
{\bf x}}^{(1),(2)}$, $\hat{{\bf p}}^{(1),(2)}$, act on $w({\bf X},\vec{\mu},%
\vec{\nu},t)$ as follows 
\begin{eqnarray}
&&\hat{x}_{k}^{(1)}=-\left( \frac{\partial }{\partial X_{k}}\right) ^{-1}%
\frac{\partial }{\partial \mu _{k}}+\frac{i}{2}\nu _{k}\frac{\partial }{%
\partial X_{k}}=\hat{x}_{k}^{(2)*}  \nonumber \\
&& \\
&&\hat{p}_{k}^{(1)}=-\frac{i}{2}\mu _{k}\frac{\partial }{\partial X_{k}}%
-\left( \frac{\partial }{\partial X_{k}}\right) ^{-1}\frac{\partial }{%
\partial \nu _{k}}=-\hat{p}_{k}^{(2)*}\,.  \nonumber
\end{eqnarray}

For the propagator $\Pi \left( X,\mu ,\nu ,X^{\prime },\mu ^{\prime },\nu
^{\prime },t\right) $ corresponding to Eq.(\ref{3}) one obtains an equation
where the quantum contributions are explicitly expressed by a series in
powers of $\hbar $. To do this, we first introduce some notation.

Let $n$-vectors ${\bf x}$ and ${\bf p}$ be described by one 2$n$-vector $%
Q_{\alpha }$, $\alpha =1,2,\ldots ,2n$, with components $(p_{1},p_{2},\ldots
,p_{n},x_{1},x_{2},\ldots ,x_{n})$, i.e., 
\[
Q_{1}=p_{1}\,,\quad Q_{2}=p_{2}\,,\quad Q_{n}=p_{n}\,,\quad
Q_{n+1}=x_{1}\,,\quad Q_{2n}=x_{n}\,
\]
Let also define the operator-vector $\widetilde{{\bf Q}}$ with components $%
\widetilde{Q}_{\alpha }$ $(\alpha =1,2,\ldots ,2n)$ 
\begin{eqnarray*}
&&\widetilde{Q}_{1}=-\left( \frac{\partial }{\partial X_{1}}\right) ^{-1}%
\frac{\partial }{\partial \nu _{1}}\,;\widetilde{Q}_{2}=-\left( \frac{%
\partial }{\partial X_{2}}\right) ^{-1}\frac{\partial }{\partial \nu _{2}}%
\,;\ldots \,;\widetilde{Q}_{n}=-\left( \frac{\partial }{\partial X_{n}}%
\right) ^{-1}\frac{\partial }{\partial \nu _{n}}\, \\
&&\widetilde{Q}_{n+1}=-\left( \frac{\partial }{\partial X_{1}}\right) ^{-1}%
\frac{\partial }{\partial \mu _{1}}\,;\widetilde{Q}_{n+2}=-\left( \frac{%
\partial }{\partial X_{2}}\right) ^{-1}\frac{\partial }{\partial \mu _{2}}%
\,;\ldots \,;\widetilde{Q}_{2n}=-\left( \frac{\partial }{\partial X_{n}}%
\right) ^{-1}\frac{\partial }{\partial \mu _{n}}\,
\end{eqnarray*}
and a 2$n$-vector ${\bf d}$ with the components 
\begin{eqnarray*}
&&d_{1}=-\frac{\mu _{1}}{2}\frac{\partial }{\partial X_{1}}\,,\quad d_{2}=-%
\frac{\mu _{2}}{2}\frac{\partial }{\partial X_{2}}\,,\quad \ldots \quad
d_{n}=-\frac{\mu _{n}}{2}\frac{\partial }{\partial X_{n}}\,, \\
&&d_{n+1}=\frac{\nu _{1}}{2}\frac{\partial }{\partial X_{1}}\,,\quad d_{n+2}=%
\frac{\nu _{2}}{2}\frac{\partial }{\partial X_{2}}\,,\quad \ldots \quad
d_{2n}=\frac{\nu _{n}}{2}\frac{\partial }{\partial X_{n}}\,.
\end{eqnarray*}
Then, the equation for the propagator of Eq.~(\ref{3}) is 
\begin{equation}
\stackrel{\bullet }{\Pi }-\frac{2}{\hbar }\sin \left( \hbar {\bf d}\frac{%
\partial }{\partial {\bf Q}}\right) H({\bf Q})|_{{\bf Q}\rightarrow 
\widetilde{{\bf Q}}}\Pi =0\,,  \label{P1}
\end{equation}
where 
\[
{\bf d}\frac{\partial }{\partial {\bf Q}}\equiv \sum_{\alpha
=1}^{2n}d_{\alpha }\frac{\partial }{\partial Q_{\alpha }}\,.
\]
Using the series expansion for $\sin \alpha $ and separating the
classical-limit term from the quantum corrections one obtains 
\begin{equation}
\stackrel{\bullet }{\Pi }-2{\bf d}\frac{\partial }{\partial {\bf Q}}H({\bf Q}%
)|_{{\bf Q}\rightarrow \widetilde{{\bf Q}}}\Pi -\frac{2}{\hbar }%
\sum_{n=1}^{\infty }\frac{(-1)^{n}}{(2n+1)!}\left( \hbar {\bf d}\frac{%
\partial }{\partial {\bf Q}}\right) ^{2n+1}H({\bf Q})|_{{\bf Q}\rightarrow 
\widetilde{{\bf Q}}}\Pi =0  \label{P3}
\end{equation}
In particular one sees that for quadratic interactions (local or nonlocal)
the quantum evolution is formally identical to the classical one. In the
configurational quantum cat we have a two-degree of freedom Hamiltonian 
\begin{equation}
H({\bf x,p},t)=H_{0}({\bf x,p})+H_{k}({\bf x,p})\sum_{n=-\infty }^{\infty
}\delta (t-n)\,,  \label{H1}
\end{equation}
where both $H_{0}$ and $H_{k}$ are quadratic forms in the position and
momentum operators. To write the functions explicitly, we define a 4-vector 
\begin{equation}
{\bf Q}=\left( 
\begin{array}{c}
{\bf p} \\ 
{\bf x}
\end{array}
\right)   \label{H2}
\end{equation}
and symmetric 4$\times $4-matrices $B_{0}$ and $B_{k}$.

Then the Hamiltonians $H_{0}$ and $H_{k}$ are taken in the form 
\begin{equation}
H_{0}=\frac{1}{2}{\bf Q}B_{0}{\bf Q}\,,\qquad H_{k}=\frac{1}{2}{\bf Q}B_{k}%
{\bf Q}  \label{H3}
\end{equation}
The system with the Hamiltonian~(\ref{H1}) has four linear integrals of
motion~\cite{MankVil2}
\begin{equation}
{\bf I}(t)=\Lambda (t){\bf Q}\,,  \label{H4}
\end{equation}
where the symplectic matrix satisfies the equation 
\begin{equation}
\dot{\Lambda}(t)=\Lambda \Sigma B(t)\,,  \label{H5}
\end{equation}
with a 4$\times $4-matrix $\Sigma $ with identity 2$\times $2-blocks 
\begin{equation}
\Sigma =\left( 
\begin{array}{cl}
0 & 1 \\ 
-1 & 0
\end{array}
\right)   \label{H6}
\end{equation}
and 
\begin{equation}
B(t)=B_{0}+B_{k}\sum_{n=-\infty }^{\infty }\delta (t-n)\,.  \label{H7}
\end{equation}
The initial condition for the 4$\times $4-matrix $\Lambda (t)$ is 
\begin{equation}
\Lambda (0)=1\,  \label{H8}
\end{equation}
The Floquet solution to (\ref{H5}) has the form 
\begin{equation}
\Lambda (1_{+})=e^{\Sigma B_{0}}\,e^{\Sigma B_{k}}  \label{H9}
\end{equation}
and for $n$ kicks 
\begin{equation}
\Lambda (n_{+})=\Lambda ^{n}(1_{+})=\left( e^{\Sigma B_{0}}~~e^{\Sigma
B_{k}}\right) ^{n}.  \label{H10}
\end{equation}
The equation for the propagator is 
\begin{eqnarray}
&&\stackrel{\bullet }{\Pi }+i\left[ H_{0}\left( \hat{{\bf x}}^{(1)},\hat{%
{\bf p}}^{(1)}\right) +H_{k}\left( \hat{{\bf x}}^{(1)},\hat{{\bf p}}%
^{(1)}\right) \sum_{n=-\infty }^{\infty }\delta (t-n)\right] \Pi 
\label{H11} \\
&&-i\left[ H_{0}\left( \hat{{\bf x}}^{(2)},-\hat{{\bf p}}^{(2)}\right)
+H_{k}\left( \hat{{\bf x}}^{(2)},-\hat{{\bf p}}^{(2)}\right) \sum_{n=-\infty
}^{\infty }\delta (t-n)\right] \Pi =0\,.
\end{eqnarray}
the vector-operators being 
\begin{eqnarray}
&&\hat{{\bf x}}^{(1)}=\left( -\left[ \frac{\partial }{\partial X_{1}}\right]
^{-1}\frac{\partial }{\partial \mu _{1}}+\frac{i}{2}\nu _{1}\frac{\partial }{%
\partial X_{1}}\,,-\left[ \frac{\partial }{\partial X_{2}}\right] ^{-1}\frac{%
\partial }{\partial \mu _{2}}+\frac{i}{2}\nu _{2}\frac{\partial }{\partial
X_{2}}\right) \,, \\
&&\hat{{\bf p}}^{(1)}=\left( -\frac{i}{2}\mu _{1}\frac{\partial }{\partial
X_{1}}-\left[ \frac{\partial }{\partial X_{1}}\right] ^{-1}\frac{\partial }{%
\partial \nu _{1}}\,,-\frac{i}{2}\mu _{2}\frac{\partial }{\partial X_{2}}%
-\left[ \frac{\partial }{\partial X_{2}}\right] ^{-1}\frac{\partial }{%
\partial \nu _{2}}\right) ,  \nonumber
\end{eqnarray}
and 
\[
\hat{{\bf x}}^{(2)}=\hat{{\bf x}}^{(1)*},\qquad \hat{{\bf p}}^{(2)}=-\hat{%
{\bf p}}^{(1)*}.
\]
The equation for the Fourier component $G({\bf k},\vec{\mu},\vec{\nu},t)$
used to compute the Lyapunov exponent is the same as Eq.(\ref{H11}) with 
\[
\frac{\partial }{\partial X_{1}},\,\frac{\partial }{\partial X_{2}}%
\rightarrow ik_{1},\,ik_{2}\,.
\]

The equation~(\ref{H11}) can be integrated by 
\begin{equation}
\Pi \left( {\bf X},\vec{\mu},\vec{\nu},n\right) =\Pi \left( {\bf X},\vec{\mu}%
_{\Lambda },\vec{\nu}_{\Lambda },0\right) ,  \label{H13}
\end{equation}
where the parameters $\vec{\mu}_{\Lambda },\vec{\nu}_{\Lambda }$ are
expressed in terms of $\vec{\mu}$ and $\vec{\nu}$ by the matrix product rule 
\begin{equation}
\left( \vec{\nu}_{\Lambda },\vec{\mu}_{\Lambda }\right) =\left( \vec{\nu},%
\vec{\mu}\right) \Lambda ^{-1}(n_{+})\,,  \label{H14}
\end{equation}
where $\Lambda ^{-1}(n_{+})$ is given by (\ref{H10}).

For the Hamiltonians $H_{1}$ and $H_{2}$ of Eqs.(\ref{qc1}) and (\ref{qc2})
the matrices $B_{0}$ and $B_{k}$ are 
\[
B_{0}=\left( 
\begin{array}{crcl}
1 & 0 & 0 & 1 \\ 
0 & 1 & 0 & 0 \\ 
0 & 0 & 0 & 0 \\ 
1 & 0 & 0 & 0
\end{array}
\right) ,\qquad B_{k}=\left( 
\begin{array}{crcl}
0 & 0 & 0 & 0 \\ 
0 & 0 & 1 & 0 \\ 
0 & 1 & 0 & 0 \\ 
0 & 0 & 0 & 0
\end{array}
\right) 
\]
for $H_{1}$ and 
\[
B_{0}=\left( 
\begin{array}{crcl}
1 & 0 & 0 & 0 \\ 
0 & 1 & 0 & 0 \\ 
0 & 0 & 0 & 0 \\ 
0 & 0 & 0 & 0
\end{array}
\right) ,\qquad B_{k}=\left( 
\begin{array}{crcl}
0 & 0 & 0 & 1 \\ 
0 & 0 & 1 & 1 \\ 
0 & 1 & 0 & 0 \\ 
1 & 1 & 0 & 0
\end{array}
\right) 
\]
for $H_{2}$.

For the model~\cite{Mendes1} where only the kick contributions are kept in
the Hamiltonian one has 
\[
B_{0}=0 
\]
and 
\[
B_{k}=\frac{\ln (1+\omega )}{\omega +2}\left( 
\begin{array}{crcl}
0 & {\cal L}(\omega ) &  &  \\ 
{\cal L}(\omega ) & 0 &  & 
\end{array}
\right) , 
\]
the 2$\times $2-matrix being 
\[
{\cal L}(\omega )=\left( 
\begin{array}{crcl}
-\omega & \frac{2(1+\omega )}{\omega } &  &  \\ 
2\omega & \omega &  & 
\end{array}
\right) , 
\]
with $\omega =(1+\sqrt{5})/2$.

For this model the matrix $\Lambda ^{-1}(n_{+})$ in (\ref{H14}) is given by 
\[
\Lambda ^{-1}(n_{+})=\left( 
\begin{array}{cr}
\widetilde{f}^{n} & 0 \\ 
0 & \widetilde{f}^{-n}
\end{array}
\right) ,
\]
where 
\[
\widetilde{f}^{n}=\left( 
\begin{array}{cr}
\omega ^{-2n+1}+\omega ^{2n-1} & -\omega ^{-2n}+\omega ^{2n} \\ 
-\omega ^{-2n}+\omega ^{2n} & \omega ^{-2n+1}+\omega ^{2n+1}
\end{array}
\right) 
\]
For all three models the equations~(\ref{H11}) are the same for classical
and quantum motion. Therefore the Lyapunov exponent must be the same in the
classical and quantum cases. In particular, as is known from the classical
case\cite{Mendes1}, there is a positive Lyapunov exponent, namely 
\[
\lambda =\ln \omega ^{2}
\]

\subsection{The standard map}

The standard map is a case where the phenomena of wave function localization
is believed to have a taming effect on chaos. The Lyapunov exponent analysis
gives a characterization of how this taming effect comes about.

The Hamiltonian is 
\begin{equation}
H=\frac{p^{2}}{2}+\gamma \cos \left( q\right) \sum_{n=-\infty }^{\infty
}\delta \left( t-n\tau \right)  \label{6.35}
\end{equation}
the configuration space being now the circle, $q\in S^{1}$. This system
describes a particle rotating in a ring and subjected to periodic kicks. It
has been extensively used in studies of quantum chaos (\cite{Izrailev} - 
\cite{Zaslavski}) and has even been tested experimentally with ultra-cold
atoms trapped in a magneto-optic trap\cite{Raizen}.

From (\ref{6.6}) the equation to be solved now is 
\begin{equation}
\frac{\partial G}{\partial t}-\mu \frac{\partial G}{\partial \nu }-\frac{%
\gamma }{\hbar }\sum_{n=-\infty }^{\infty }\delta \left( t-n\tau \right)
\sin \left( \frac{\hbar }{2}\nu \right) \left\{ G\left( 1,\mu +1,\nu
,t\right) -G\left( 1,\mu -1,\nu ,t\right) \right\} =0  \label{6.36}
\end{equation}
where we have specialized to the value $k=1$ because this is the only $k-$%
value needed to compute the Lyapunov exponent (\ref{6.5}). Notice that we
have used here the same tomographic transformations that were described in
Section 2 for functions on the line. This is justified by considering all
functions as defined not in $S^{1}$ but in the suspension of $S^{1}$.

From (\ref{6.36}) one sees that between any two kicks the function
propagates freely, namely 
\begin{equation}
G\left( 1,\mu ,\nu ,t_{0}\right) \rightarrow G\left( 1,\mu ,\nu
,t_{1_{-}}\right) =G\left( 1,\mu ,\nu +\mu \tau ,t_{0}\right)  \label{6.37}
\end{equation}
and at the time of the kick a quantity is added that is proportional to a
finite difference (in $\mu $). 
\begin{equation}
G\left( 1,\mu ,\nu ,t_{1_{+}}\right) =G\left( 1,\mu ,\nu ,t_{1_{-}}\right) +%
\frac{\gamma }{2}f\left( \nu \right) \left\{ G\left( 1,\mu +1,\nu
,t_{1_{-}}\right) -G\left( 1,\mu -1,\nu ,t_{1_{-}}\right) \right\}
\label{6.38}
\end{equation}
where, for the classical case 
\begin{equation}
f\left( \nu \right) =\nu  \label{6.39}
\end{equation}
and for the quantum case 
\begin{equation}
f\left( \nu \right) =\frac{2}{\hbar }\sin \left( \frac{\hbar }{2}\nu \right)
\label{6.40}
\end{equation}

To compute the Lyapunov exponent we need the evolution of the derivatives
(in $\mu $ and $\nu $) of $G$ at $\mu =\nu =0$. From (\ref{6.37}) and (\ref
{6.38}) one obtains the following iteration for the derivatives 
\begin{equation}
\begin{array}{lll}
G^{(2)}\left( 1,0,0,t+1\right) & = & G^{(2)}\left( 1,0,0,t\right)
+G^{(3)}\left( 1,0,0,t\right) \\ 
G^{(3)}\left( 1,0,0,t+1\right) & = & G^{(2)}\left( 1,0,0,t\right) +\frac{%
\gamma }{2}\left( G\left( 1,1,\tau ,t\right) -G\left( 1,-1,-\tau ,t\right)
\right)
\end{array}
\label{6.41}
\end{equation}

Let us consider first the classical case ($\hbar =0$ , $f\left( \nu \right)
=\nu $ and $\gamma >0$). Let also $\tau =1$ and $q_{0}=p_{0}=0$ in the
initial condition (\ref{6.7}). Then, one obtains the following recursion for
the derivatives of $G$ at $\mu =\nu =0$. 
\begin{equation}
\begin{array}{lll}
G^{(2)}\left( 1,0,0,n+1\right) & = & G^{(2)}\left( 1,0,0,n\right)
+G^{(3)}\left( 1,0,0,n\right) \\ 
G^{(3)}\left( 1,0,0,n+1\right) & = & \gamma G^{(2)}\left( 1,0,0,n\right)
+\left( 1+\gamma \right) G^{(3)}\left( 1,0,0,n\right)
\end{array}
\label{6.45}
\end{equation}
which has the solution 
\begin{equation}
\begin{array}{lll}
G^{(2)}\left( 1,0,0,n\right) & = & A_{n}(z)v_{1}+B_{n}(z)v_{2} \\ 
G^{(3)}\left( 1,0,0,n\right) & = & C_{n}(z)v_{1}+D_{n}(z)v_{2}
\end{array}
\label{6.46}
\end{equation}
with $z=2+\gamma $ and 
\begin{equation}
\begin{array}{lll}
A_{n}(z) & = & U_{n-1}(\frac{z}{2})-U_{n-2}(\frac{z}{2}) \\ 
B_{n}(z) & = & \frac{1}{z-2}C_{n}(z) \\ 
C_{n}(z) & = & U_{n}(\frac{z}{2})-2U_{n-1}(\frac{z}{2})+U_{n-2}(\frac{z}{2})
\\ 
D_{n}(z) & = & U_{n}(\frac{z}{2})-U_{n-1}(\frac{z}{2})
\end{array}
\label{6.47}
\end{equation}
where $U_{n}(z)=\frac{\sin \left( \left( n+1\right) \cos ^{-1}z\right) }{%
\sin \left( \cos ^{-1}z\right) }$ is a Chebyshev polynomial.

For the Lyapunov exponent one obtains in this case 
\begin{equation}
\lambda =\ln \left| 1+\frac{1}{2}\gamma +\sqrt{\frac{1}{4}\gamma ^{2}+\gamma 
}\right|  \label{6.48}
\end{equation}
a result similar to the harmonic kicks on the line. One sees that as long as 
$\gamma >0$ the exponent $\lambda $ in Eq.(\ref{6.48}) is always positive.
This results from the choice made for the phase space point ($p_{0}=q_{0}=0$%
) where the marginal distribution receives the singular perturbation (\ref
{L9}). If instead we had chosen ($p_{0}=0$ and $q_{0}=\pi $) in the initial
condition (\ref{6.7}), one sees easily by a change of coordinates in the
Hamiltonian that this is equivalent to replace $\gamma $ by $-\gamma $. Then
the Lyapunov exponent $\lambda $ in Eq.(\ref{6.48}) is positive only for $%
\gamma >4$. As discussed at length in the next section, this only means that
it is the phase space point ($p_{0},q_{0}$) that defines the measure for
which the Lyapunov exponent is computed. Hence, for the measure that
supports the hyperbolic point ($p_{0}=0$ , $q_{0}=0$) the exponent is always
positive, whereas for sufficiently small $\gamma >0$ the exponent for the
measure that supports the elliptic point ($p_{0}=0$ , $q_{0}=\pi $) is
negative.

For the quantum case ($\hbar \neq 0$) let us consider an initial condition $%
G\left( 1,\mu ,\nu ,0\right) =\mu +\nu $ (corresponding to $p_{0}=0$ , $%
q_{0}=0$, $v_{1}=v_{2}=1$) and $\tau =1$. According to Eq.(\ref{6.41}), all
one needs to compute the Lyapunov exponent is the time evolution of $G\left(
1,1,1,t\right) $. For this purpose we set up a matrix recursion for the
evolution equations (\ref{6.37}-\ref{6.38}). Define the following matrices 
\begin{equation}
M_{0}=\left( 
\begin{array}{lll}
1 & 1 & 0 \\ 
0 & 1 & 0 \\ 
0 & 0 & 1
\end{array}
\right) ;M_{+}=\left( 
\begin{array}{lll}
1 & 1 & 0 \\ 
0 & 1 & 0 \\ 
1 & 1 & 1
\end{array}
\right) ;M_{-}=\left( 
\begin{array}{rrr}
1 & 1 & 0 \\ 
0 & 1 & 0 \\ 
-1 & -1 & -1
\end{array}
\right)  \label{6.49}
\end{equation}
and vectors $\left( 
\begin{array}{l}
\alpha \\ 
\beta \\ 
\gamma
\end{array}
\right) $ where $\alpha $ counts the number of $\mu $'s, $\beta $ the number
of $\nu $'s and $\gamma $ is a simple number. Then with $Tr$ denoting the
sum of the elements in a vector and 
\begin{equation}
x_{0}=\left( 
\begin{array}{l}
1 \\ 
1 \\ 
0
\end{array}
\right) ;y_{0}=\left( 
\begin{array}{l}
0 \\ 
1 \\ 
0
\end{array}
\right)  \label{6.50}
\end{equation}
the initial condition is $G\left( 1,1,\tau ,0\right) =Tr(x_{0})$ and the
function $f\left( \nu \right) =f\left( Tr(y_{0})\right) $.

On arbitrary functions of 3-dimensional vectors, the operators $%
K_{0},K_{+},K_{-}$ act on the arguments by the matrices $M_{0},M_{+},M_{-}$%
\begin{equation}
K_{i}g\left( x\right) =g\left( M_{i}x\right)  \label{6.51}
\end{equation}
Then 
\begin{equation}
G\left( 1,1,1,n\right) =Tr\left\{ \left( K_{0}+\frac{\gamma }{2}%
f(y_{0})\left[ K_{+}-K_{-}\right] \right) ^{n}x_{0}\right\}  \label{6.52}
\end{equation}
where it is understood that the power of the operator is fully expanded
before the $Tr$ operation is applied to each one of the vector arguments.
When this expansion is made, one obtains an expression of the form 
\begin{equation}
G\left( 1,1,1,n\right) =n+2+Tr\left\{ \sum_{k=1}^{n}\left( \frac{\gamma }{2}%
\right) ^{k}\sum_{i=1}^{\frac{2^{k-1}n!}{k!\left( n-k\right) !}}c_{i}f\left(
\prod M_{i_{1}}y_{0}\right) \cdots f\left( \prod M_{i_{k}}y_{0}\right)
\right\}  \label{6.53}
\end{equation}
The products of $M$ matrices in the arguments of $f\left( \cdot \right) $
contain a variable number of factors, from $1$ to $k$. However for each
term, a different combination of products will appear. For $\hbar \neq 0$
the function $f\left( \nu \right) $ is proportional to a sine and, if $\frac{%
\hbar \tau }{4\pi }$ is irrational, the coefficient of each $\gamma ^{k}$
behaves like a sum of random variables of zero mean. Therefore each
coefficient averages to zero and 
\begin{equation}
G\left( 1,1,1,n\right) \sim n+2  \label{6.54}
\end{equation}
Large fluctuations are however to be expected in view of the large number of
terms in the sums for large $n$. From (\ref{6.41}), the result (\ref{6.54})
now implies 
\[
\begin{array}{lll}
G^{(2)}\left( 1,0,0,n\right) & \sim & 1+n\left( 1-\gamma \right) +\frac{%
\gamma }{2}\left( \frac{n\left( n+1\right) }{2}+\frac{n\left( n+1\right)
\left( 2n+1\right) }{6}\right) \\ 
G^{(3)}\left( 1,0,0,n\right) & \sim & 1+\gamma \left( \frac{\left(
n+1\right) \left( n+2\right) }{2}-1\right)
\end{array}
\]
For large $n$, $\log G^{(2)}\left( 1,0,0,n\right) \sim 3\log n$ and $\log
G^{(2)}\left( 1,0,0,n\right) \sim 2\log n$ and the Lyapunov exponent
vanishes.

The situation we have been studying ($p_{0}=q_{0}=0$ in the initial
perturbation) corresponds to the (hyperbolic) case where the classical
Lyapunov exponent is positive for any $\gamma $. We see here clearly the
taming effect of quantum mechanics on classical chaos and its dynamical
origin. It results from the replacement in the evolution equation of the
linear function $f\left( \nu \right) =\nu $ by $f\left( \nu \right) =\frac{2%
}{\hbar }\sin \left( \frac{\hbar }{2}\nu \right) $. This in turn is a
consequence of the replacement of the classical Boltzman equation by the
quantum evolution equation (\ref{defeq}), or in algebraic terms, by the
replacement of the ordinary product by the Moyal-Vey product in the
non-commutative quantum phase space.


In this model, the origin of the taming effect of quantum mechanics on
classical chaos, is traced back to the existence, in the $\hbar -$deformed
equation (\ref{6.6}), of infinitely many terms in the series which add up to
a bounded function in $\nu $. How general this mechanism is, for other
quantum systems, is an open question. In any case the taming effect of
quantum mechanics, obtained here for the standard map, is more accurate than
previous discussions of the same system, because it refers to the behavior
of the Lyapunov exponent rather than to indirect chaos symptoms, like the
energy growth or diffusion behavior.

\section{Remarks and conclusions}

1- The method developed in this paper, for the quantum Lyapunov exponents,
provides a fairly unambiguous construction of these quantities, in the sense
that classical and quantum exponents have the same functional form. The
difference lies only on the time-evolution laws for the propagators.

The dynamical evolution laws of the marginal distributions, obtained by the
tomographic map, are apparently more complex that the familiar
Schr\"{o}dinger equation. However, for the computation of the Lyapunov
exponents, they provide a fairly efficient computational scheme.

Quantum mechanics is widely believed to have a taming effect on classical
chaos. However, most discussions are of a qualitatively nature and fail to
identify the conditions under which the taming effect is expected to occur
and those in which it will not occur. This is a very relevant question in
view of the fact that for local quadratic potentials, the quantum behavior
differs very little from the classical one and genuine examples of quantum
chaos with bounded configuration space are known, like the four-dimensional
or configurational quantum cat \cite{Mendes1} \cite{Weigert}.

In the standard map, studied in Sect.5, it is clear that the suppression of
chaos is directly related both to the nature of the potential and the
analytical structure of the series in Eq.(\ref{6.6}). This operational
series, when acting on the potential, convert an unbounded function into a
bounded function in $\nu $ (the symplectic parameter conjugate to $p$). The
structure of the series corresponds to the structure of the Moyal bracket
and the way the quadratic potential (and presumably other polynomial
potentials) avoid the suppression effect, is by truncating the action of the
Moyal bracket to a finite number of cocycles.

The fact that, for non-polynomial interactions, all derivatives of the
potential intervene in the quantum evolution, means, by an analyticity
argument, that the future evolution of any local perturbation depends
strongly on what is going on at all other points. This interference between
quantum ''trajectories'' is probably the decisive factor that determines the
nature of the quantum modifications of classical chaos.

2 - A second important question concerns the support properties of the
Lyapunov exponents that have been constructed. In classical mechanics,
Lyapunov exponents are ergodic invariants. That means that they are defined
in the support of some measure. In the construction (both classical and
quantum) developed in Sect.3, the Lyapunov is obtained from a singular
perturbation of the $X-$coordinate at the point $\mu q_{0}+\nu p_{0}$ for
each pair $\left( \mu ,\nu \right) $. For the classical case, the
interpretation is clear. From the point of view of measures in phase-space,
it means that one is constructing the Lyapunov exponent that corresponds to
the measure whose support contains the point $\left( q_{0},p_{0}\right) $.

Measures on classical phase-space may be interpreted as measures on the
joint spectrum of the (commuting) operators $q$ and $p$. Therefore in the
classical case a measure $\mu $ plays the double role of a probability
measure in phase-space and a spectral measure for the dynamical operators.
For the quantum case, however, $q$ and $p$ do not commute and there is no
joint spectrum for these operators. Then, instead of one measure playing a
double role we have two:

- One is the state that is perturbed. This is the analog of the classical
probability measure, because states are the non-commutative analogs of Borel
measures.

- The other is the spectral measure of the operator $X$, the perturbation
acting, for each pair $\left( q_{0},p_{0}\right) $, at the point $\mu
q_{0}+\nu p_{0}$ of the spectrum.

In conclusion: the interpretation of the quantum Lyapunov exponent as an
ergodic invariant requires two measures, a state and a spectral measure. In
the classical case the two measures coincide.






\begin{thebibliography}{99}
\bibitem{Connes}  A. Connes, H. Narnhofer and W. Thirring; Commun. Math.
Phys. 112 (1987) 691.

\bibitem{Lindblad}  G. Lindblad; {\it Dynamical entropy for quantum systems}%
, in {\it Quantum probability and applications}, Lecture Notes in
Mathematics 1303, page 183, Springer, Berlin 1988.

\bibitem{Narnhofer2}  H. Narnhofer and W. Thirring; Commun. Math. Phys. 125
(1989) 564.

\bibitem{Gutzwiller}  M. Gutzwiller; {\it Chaos in classical and quantum
mechanics}, Springer, Berlin 1990.

\bibitem{Narnhofer1}  H. Narnhofer; J. Math. Phys. 33 (1992) 1502.

\bibitem{Haake}  F. Haake, H. Wiedemann and W. Zyczkowski; Ann. Physik 1
(1992) 531.

\bibitem{Mendes1}  R. Vilela Mendes; Phys. Lett. A{\bf 171} (1992) 253.

\bibitem{Majewski1}  W. A. Majewski and M. Kuna; J. Math. Phys. 34 (1993)
5007.

\bibitem{Ingraham}  R. L. Ingraham and G. L. Acosta; Phys. Lett. A181 (1993)
450.

\bibitem{Benatti2}  F. Benatti; {\it Deterministic chaos in infinite quantum
systems}, Springer, Berlin 1993.

\bibitem{Vilela4}  R. Vilela Mendes; Phys. Lett. A187 (1994) 299.

\bibitem{Vilela2}  R. Vilela Mendes; {\it Entropy and quantum characteristic
exponents. Steps towards a quantum Pesin theory, }in{\it \ Chaos - The
interplay between stochastic and deterministic behavior, }P. Garbaczewski,
M. Wolf and A. Weron (Eds.), page 273, Lecture Notes in Physics 457,
Springer, Berlin 1995.

\bibitem{Roepstorff}  G. Roepstorff; {\it Quantum dynamical entropy} in {\it %
Chaos - The interplay between stochastic and deterministic behavior, }P.
Garbaczewski, M. Wolf and A. Weron (Eds.), page 305, Lecture Notes in
Physics 457, Springer, Berlin 1995.

\bibitem{Majewski2}  W. A. Majewski; {\it Applications of quantum
characteristic exponents }in {\it Chaos - The interplay between stochastic
and deterministic behavior, }P. Garbaczewski, M. Wolf and A. Weron (Eds.),
page 273, Lecture Notes in Physics 507, Springer, Berlin 1995.

\bibitem{Alicki}  P. Alicki, D. Makowiec and W. Miklaszewski; Phys. Rev.
Lett. 77 (1996) 838.

\bibitem{Faisal}  F. H. M. Faisal and U. Schwengelbeck; Phys. Lett. A207
(1995) 31.

\bibitem{Schack}  R. Schack and C. M. Caves; Phys. Rev. E53 (1996) 3257.

\bibitem{Pettini}  G. Iacomelli and M. Pettini; Phys. Lett. A212 (1996) 29.

\bibitem{Knauf}  A. Knauf and Y. G. Sinai; {\it Classical nonintegrability,
quantum chaos}, Birkh\"{a}user, Basel 1997.

\bibitem{Bunakov}  V. E. Bunakov, F. F. Valiev and Yu. M. Tchuvilski; Phys.
Lett. A243 (1998) 288.

\bibitem{Mendes2}  R. Vilela Mendes and R. Coutinho; Phys. Lett. A{\bf 239},
(1998) 239.

\bibitem{Benatti1}  F. Benatti and M. Fannes; J. Phys. A: Math. Gen. 31
(1998) 9123.

\bibitem{Majewski3}  W. A. Majewski; Chaos, Solitons and Fractals 9 (1998)
77.

\bibitem{Brumer}  A. K. Pattanayak and P. Brumer; Phys. Rev. E56 (1997) 5174.

\bibitem{Garcia98}  G. G. de Polavieja, Phys. Rev. A{\bf 57} (1998) 3184.

\bibitem{Winter}  C. van Winter; J. Math. Phys. 40 (1999) 123.

\bibitem{Lichne1}  F. Bayen, M. Flato, M. Fronsdal, A. Lichnerowicz and D.
Sternheimer, Ann. Phys. 111 (1977) 61 - 151.

\bibitem{Wigner32}  E. Wigner, Phys. Rev. {\bf 40} (1932) 749.

\bibitem{Mancini95}  S. Mancini, V. I. Man'ko, and P. Tombesi, Quantum
Semiclass. Opt. {\bf 7} (1995) 615.

\bibitem{Dariano96}  G. M. D'Ariano, S. Mancini, V. I. Man'ko, and P.
Tombesi, Quantum Semiclass. Opt. {\bf 8} (1996) 1017.

\bibitem{ManciniPL}  S. Mancini, V. I. Man'ko, and P. Tombesi, Phys. Lett. A%
{\bf 213} (1996) 1.

\bibitem{ManciniFP}  S. Mancini, V. I. Man'ko, and P. Tombesi, Found. Phys. 
{\bf 27} (1997) 801.

\bibitem{OlgaJRLR97}  Olga Man'ko and V. I. Man'ko, J. Russ. Laser Research 
{\bf 18} (1997) 407.

\bibitem{Olga2}  Olga Man'ko and V. I. Man'ko; J. Russ. Laser Research {\bf %
20} (1999) 67.

\bibitem{MendesManko}  V. I. Man'ko and R. Vilela Mendes; Phys. Lett. A263
(1999) 53.

\bibitem{elaf98}  V. I. Man'ko, {\it Conventional quantum mechanics without
wave function and density matrix}, E-print quant-ph/9902079, ~in {\it %
Proceedings of the XXXI Latin American School of Physics} (XXXI ELAF,
July-August 1998, Mexico) S. Hacyan, R. J\'{a}uregui, and R. Lopez-Pe\~{n}a
(Eds.), , AIP Conference Proceedings, American Institute of Physics, vol.
464, pp. 191, New York 1999.

\bibitem{Moyal49}  J. E. Moyal, Proc. Cambridge Philos. Soc. {\bf 45} (1949)
99.

\bibitem{Mane}  R. Ma\~{n}e; {\it Ergodic theory and differentiable dynamics}%
, Springer, Berlin 1983.

\bibitem{Oseledec}  V. I. Oseledec; Trans. Moscow Math. Soc. 19 (1968) 197.

\bibitem{Raghu}  M. S. Raghunatan; Israel Jour. Math. 32 (1979) 356.

\bibitem{Vilela3}  R. Vilela Mendes; J. Phys. A: Math. Gen. 24 (1991) 4349.

\bibitem{Slawny}  G. A. Hagedorn, M. Loss and J. Slawny; J. Phys. A19 (1986)
521.

\bibitem{Weigert}  S. Weigert; Z. Phys. B80 (1990) 3.

\bibitem{Arnold}  V. I. Arnold and A. Avez; {\it Ergodic problems of
classical mechanics}, W. A. Benjamim, New York 1968.

\bibitem{MankVil2}  V. I. Man'ko and R. Vilela Mendes; Physica Scripta 56
(1997) 417.

\bibitem{Izrailev}  F. M. Izrailev and D. L. Shepelyanski; Theor. Math.
Phys. 43 (1980) 553.

\bibitem{Shepelyanski}  D. L. Shepelyanski; Physica D8 (1983) 208.

\bibitem{Casati1}  G. Casati, B. V. Chirikov, I. Guarnieri and D. L.
Shepelyanski; Phys. Rev. Lett. 56 (1986) 2437.

\bibitem{Casati2}  G. Casati; Chaos 6 (1996) 391.

\bibitem{Zaslavski}  M. Zaslavski; Chaos 6 (1996) 184.

\bibitem{Raizen}  M. G. Raizen; Comments At. Mol. Phys. 34 (1999) 321.
\end{thebibliography}
\end{document}